\documentclass[twocolumn, showpacs, amsmath, amssymb, aps, pra]{revtex4-1}
\usepackage{graphicx}
\usepackage{dcolumn}
\usepackage{bm}

\begin{document}
\title{Casimir EMF in configurations with shifted elements}

\author{Evgeny\, G.\, Fateev}
 \email{e.g.fateev@gmail.com}
\affiliation{%
Institute of mechanics, Ural Branch of the RAS, Izhevsk 426067, Russia
}%
\date{\today}

\begin{abstract}
The possibility in principle is shown for the existence of Casimir 
electromotive force (EMF) in a configuration with parallel nanosized metal 
plates which are shifted relative one another. The effect is theoretically 
demonstrated for a configuration with two plates (wings) of finite length, 
the particular case of which is classical Casimir configuration with 
parallel plates. It is found that when the plates are strictly parallel, EMF 
does not appear. However, when the plates are shifted relative one another, 
in each of them time-constant EMF is generated. It is also found 
that maximal EMF values depending on the plate shifts are larger than those 
depending on the values of angles between the wings. All the found effects 
exist in periodic configurations with shifted elements. There are optima of 
the configuration geometrical parameters at which the EMF\textbf{ 
}generation can be maximal.
\end{abstract}

\maketitle

\textbf{INTRODUCTION}

Recently the possibility in principle has been shown for the Casimir 
EMF existence in well-conducting nanosized individual 
\cite{Fateev:2015} and periodic \cite{Fateev:2016} configurations 
which are not closed in circuit. The effect is theoretically demonstrated 
with the use of configurations consisting of two nonparallel plates (wings). 
Naturally, in parallel metal plates in the classical configuration 
considered by Casimir \cite{Casimir:1948, Casimir:1949}, 
\cite{Milton:2001, Klimchitskaya:2009, Bordag:2009} no 
electromotive forces should appear. However, there can be fluctuations of 
electric potentials at the ends of the plates due to Johnson-Nyquist thermal 
noise \cite{Bimonte:2008} and interference currents because of radio 
interference.

The possibility of the Casimir EMF generation is associated with 
the effect which is similar to the light-induced electron drag which can 
appear in metals \cite{Gurevich:1992}, \cite{Shalaev:1992, 
Shalaev:1996}, graphite nano-films \cite{Mikheev:2012, 
Mikheev:2017} and semiconductors \cite{perovich:1981}. In our 
case, the drag effect can occur in nonparallel and well-conducting wings due 
to the resultant uncompensated action of virtual photons on electrons. 
Earlier it has been shown that in principle, in addition to the excitation 
of an EMF, the system with nonparallel wings can experience Casimir 
expulsion force \cite{Fateev:2012} and have some other interesting 
effects \cite{Fateev:2013, Fateev:2014}. The uncompensated action 
of forces on nonparallel configurations is due to the nonuniform action of 
Casimir forces on the opposite ends of the configuration asymmetrical along 
one of the coordinates. Both for the expulsion forces and for the EMF, the 
optimal parameters of the angles of nonparallel-plates opening and their 
lengths have been found at which the effects achieved should be maximal. In 
particular, when the plates in the configuration are shifted relative to one 
another, the forces of Casimir expulsion are not compensated as well 
\cite{Fateev:2014}. As the result, at some relative shifts, an 
increase in the expulsion forces of the entire configuration takes place, 
and torque moments and relaxation oscillation effects appear.

The question concerning the possibility and the properties of the 
EMF generation by nanosized metal configurations when their 
elements are shifted is quite interesting.

\textbf{THEORY}

Let us consider a configuration with nonparallel and shifted wings using it 
as an example for investigating the possibility of the existence of Casimir 
EMF. Let us note that each individual configuration presents by itself a 
figure comprised by flat metal plates (wings). The inner and outer surfaces 
of the figure should have the properties of almost perfect mirrors with the 
reflection coefficient $\rho $.
\begin{figure*}[htbp]
\hypertarget{fig1}
\centerline{
\includegraphics[width=2.6in,height=2.6in]{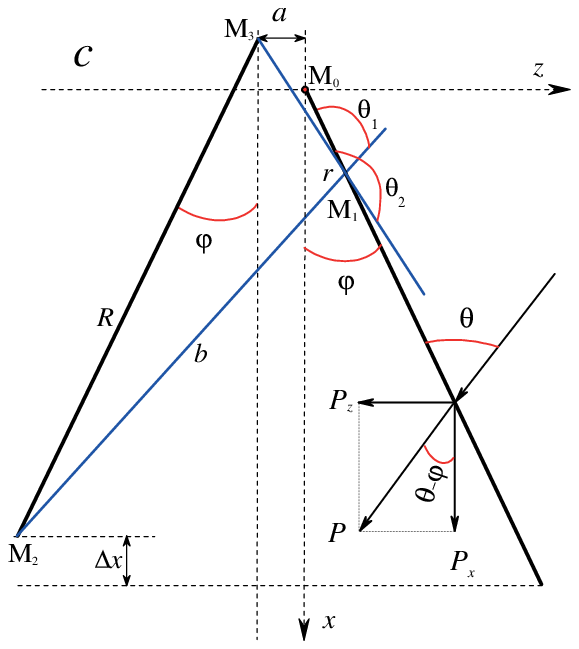}
\includegraphics[width=2.6in,height=2.6in]{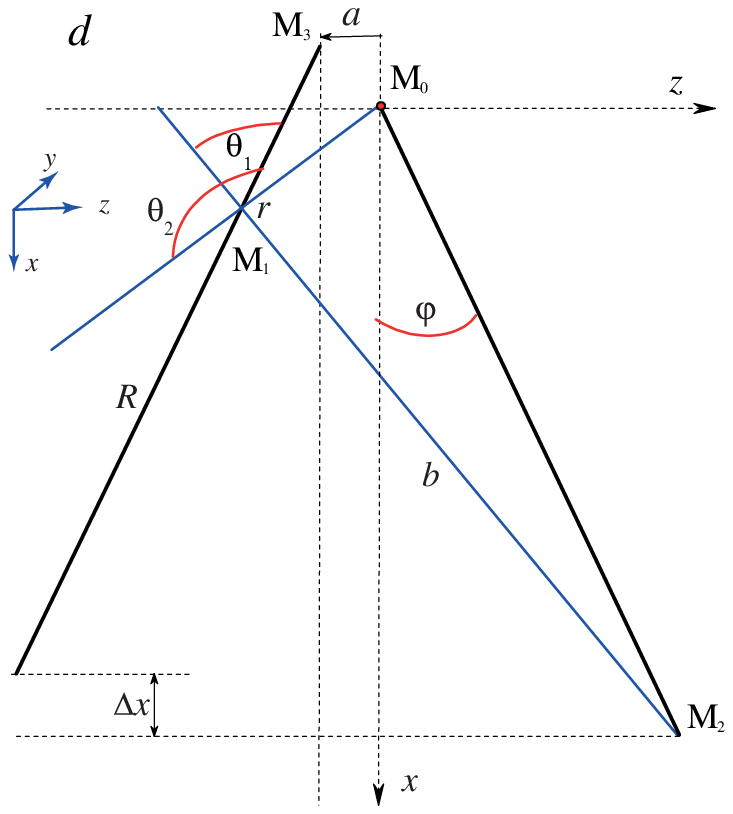}}
\label{fig1}
\caption{($c$) -- the schematic view of a symmetrical configuration with a 
shifted left wing (length $R)$ by step $\Delta x$ against the $x$-axes 
direction, the particular case of which is parallel wings (for $\varphi =0)$ 
and a triangle (at $a=0$ and $\Delta x=0)$. The figure with the width $L$ in 
the $y$-axis direction (perpendicular to the plane of Figure $L$) is shown 
in the system of Cartesian coordinates $(x,z)$ Blue straight lines designate 
the virtual rays with the length $b$outgoing from the point $\mbox{M}_1 $ at 
the limit angles $\Theta _1 $ and $\Theta _2 $ toward the figure right 
surface and finishing at the ends of the opposite wing of the figure at the 
points $\mbox{M}_2 $ and $\mbox{M}_3 $, respectively; ($d)$ -- the schematic 
view of the rays passage at the limit angles $\Theta _1 $and $\Theta _2 $ 
from an arbitrary point $\mbox{M}_1 $ of the left wing towards the ends of 
the right wing.}
\end{figure*}
\begin{figure*}[htbp]
\hypertarget{fig2}
\centerline{
\includegraphics[width=2.5in,height=2.5in]{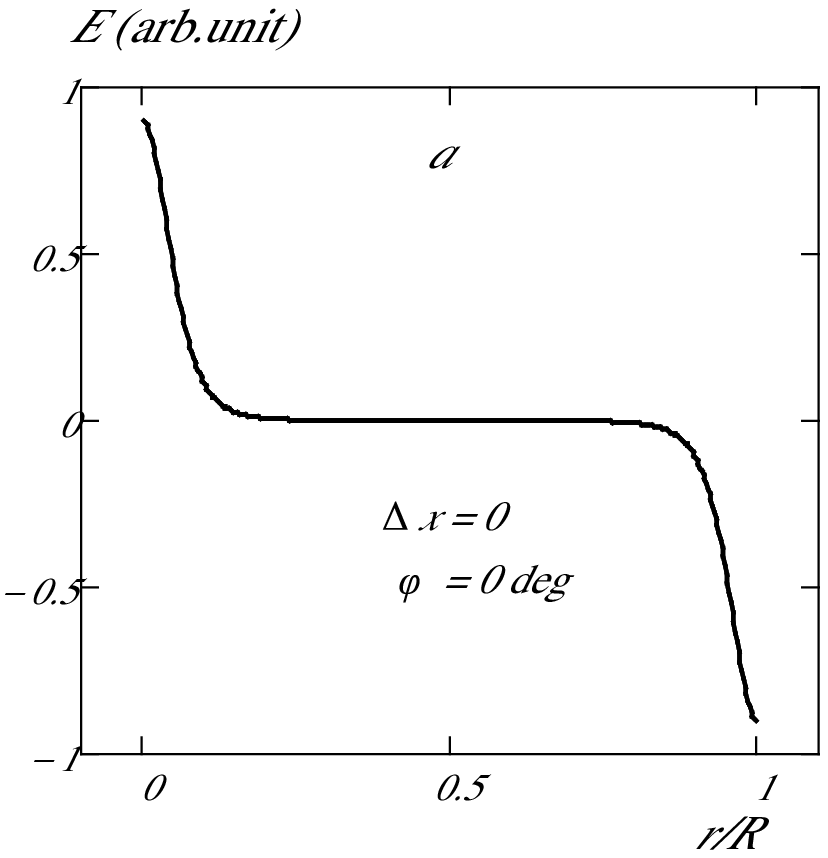}
\includegraphics[width=2.5in,height=2.5in]{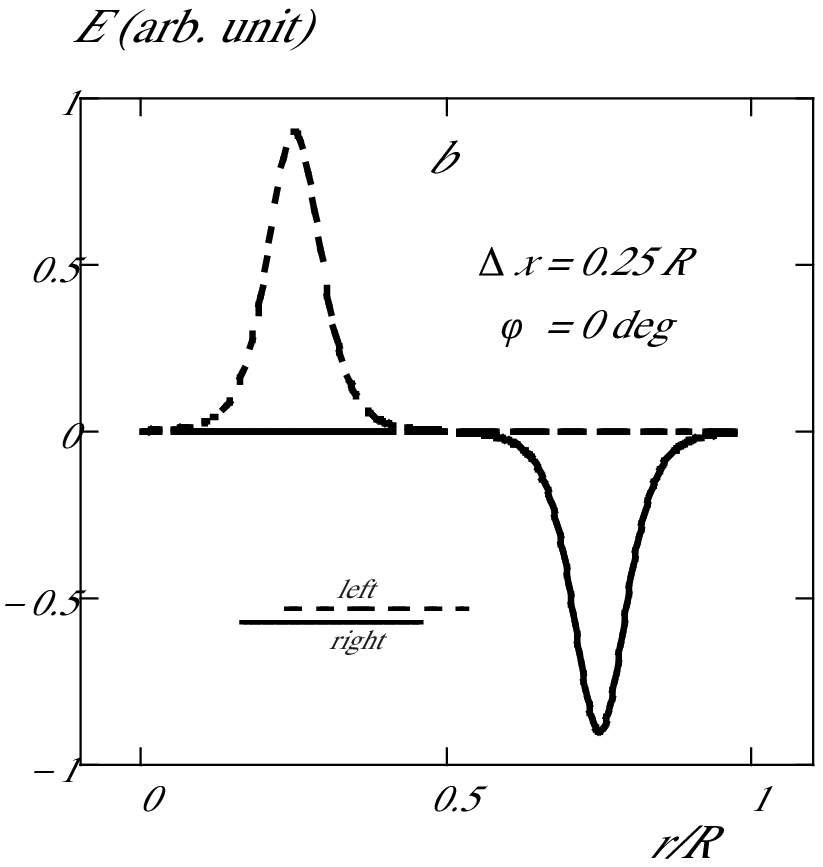}\linebreak }
\centerline{
\includegraphics[width=2.5in,height=2.5in]{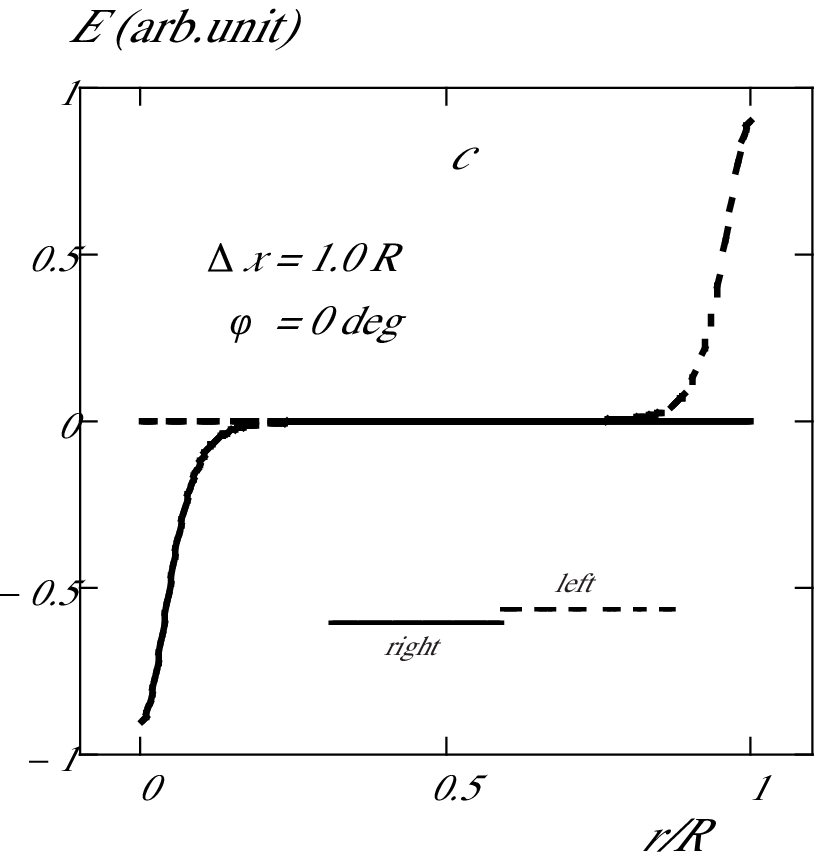}
\includegraphics[width=2.5in,height=2.5in]{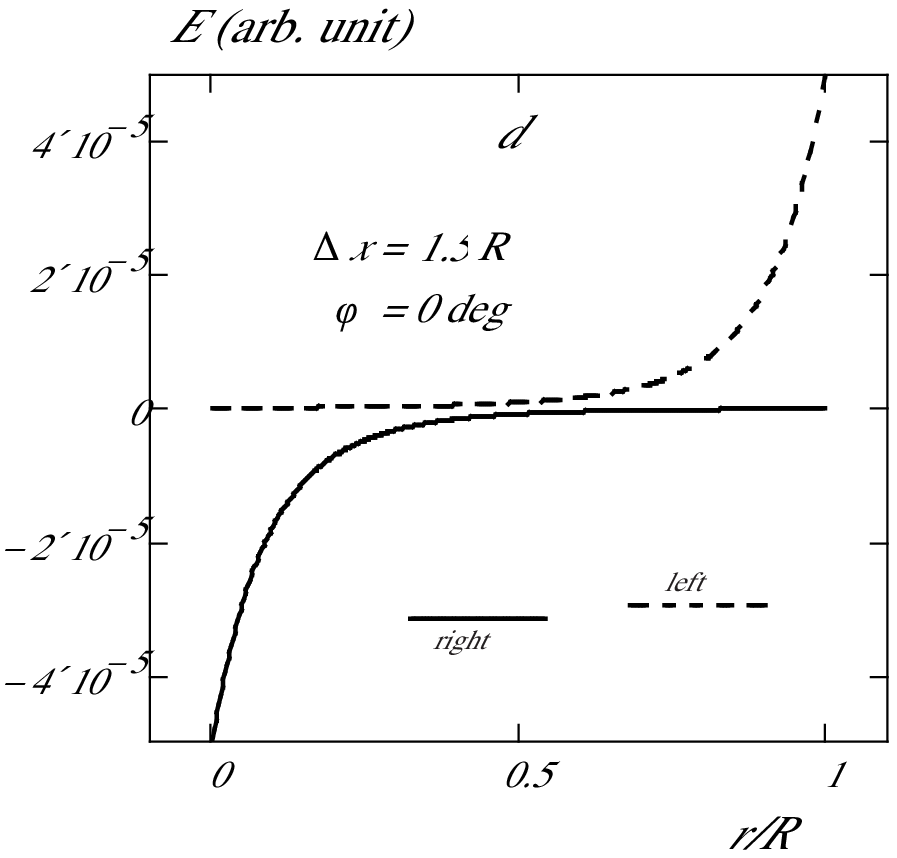}\linebreak }
\label{fig2}
\caption{Local intensities of the electric field $E$ generated along the 
parallel wings ($\varphi =0)$ at four different values of the relative shift 
$\Delta x/R$ for the same value of the distance $a$. }
\end{figure*}
Let us present an individual configuration in the system of Cartesian 
coordinates in the form of two thin metal plates with the width $L$ 
(oriented along the $z$-axis) and surface length $R$ (a wing) located at the 
distance $a$ from one another; the opening angle $\varphi $ between the plates 
can be varied (by the same value for both wings simultaneously) as shown in 
Fig.\hyperlink{fig1} 1. Also let us take into consideration the possibility of the shift of 
one of the figure wings by the step $\Delta x$ in the direction opposite to 
the $x$-axis. For such geometry, the angle $\varphi $ should not be larger than 
the value $\varphi \le \mbox{arccot}\left( {{\Delta x} \mathord{\left/ 
{\vphantom {{\Delta x} a}} \right. \kern-\nulldelimiterspace} a} \right)$, 
otherwise the situation arises when the configuration of plates with two 
neighboring figures appears. This situation requires a different problem 
statement.

Further, let us note that for the above figures the concept developed in 
Ref. \cite{Fateev:2015} is completely applicable. The EMF for 
one wing of the figure $\Delta E_\parallel $ can be found in the first 
approximation in the following form \cite{Fateev:2015}
\begin{equation}
\label{eq1}
\Delta E_\parallel =\frac{1}{2n_{b} e}\left[ {1-\rho -k} \right]\int_0^L {dy} 
\int_0^{r_{\max } } {P(\Theta ,r,\varphi )dr} 
\end{equation}
Here, $n_{b} $ - volume density, $e$ - electron charge, $\rho $- coefficient 
of reflection, $k$ - photon-transmission factor, $P(\Theta ,r,\varphi )$ - 
local specific pressure at each point $\rho $ on the wing of the figure with 
the length $r_{\max } $ and width $L$
\begin{widetext}
\begin{equation}
\label{eq2}
 P(\Theta ,r,\varphi )=\frac{\hbar c\pi ^2}{240 s^4}\int_{\Theta _1 }^{\Theta 
_2 } {d\Theta } \sin (\Theta -2\varphi )^4\sin \Theta \cos \Theta \\ 
 =-\frac{\hbar c\pi ^2}{240 s^4}A(\varphi ,\Theta _1 ,\Theta _2 ) ,\\ 
\end{equation}
where
\begin{equation}
\label{eq3}
\begin{array}{r}
 A(\varphi ,\Theta _1 ,\Theta_2)      =\frac{1}{96}\left[ {24\Theta _1 \sin 4\varphi 
 -24\Theta _2 \sin 4\varphi } \right. \\ 
 +18\cos 2\Theta _1 -18\cos 2\Theta _2 \\ 
  +6\cos (4\varphi -4\Theta _2 )-6\cos (4\varphi -4\Theta _1 ) \\ 
+3\cos (8\varphi -2\Theta _2 )-3\cos (8\varphi -2\Theta _1 ) \\ 
 \left. {+\cos (8\varphi -6\Theta _1 )-\cos (8\varphi -6\Theta _2 )} \right]. \\ 
\end{array}
\end{equation}
\end{widetext}
In formula (\ref{eq2}), $\hbar =h/2\pi $ - reduced Planck constant, and $c$- light 
speed; the functional expressions for the figure limit angles $\Theta 
_{_1}^{\mbox{right}} ,\;\Theta _{_2 }^{\mbox{right}}$  for the right 
wing in the configuration taking into account relative shifts $\Delta x$ 
have the following form [16]

\begin{widetext}
\begin{equation}
\label{eq4}
\Theta_{1}^{\mbox{right}} =\mbox{arcos}\left\{ {-\frac{r+a\sin \varphi 
+\Delta x\cos \varphi -R\cos 2\varphi }{\sqrt {\left[ {a+(R+r)\sin \varphi } 
\right]^2+\left[ {\Delta x+(r-R)\cos \varphi } \right]^2} }} \right\},
\end{equation}
\begin{equation}
\label{eq5}
\Theta _{_2 }^{\mbox{right}} =\mbox{arccos}\left\{ {-\frac{r+a\sin \varphi 
+\Delta x\cos \varphi }{\sqrt {a^2+\Delta x^2+r^2+2a\,r\sin \varphi +2\Delta 
x{\kern 1pt}r\cos \varphi } }} \right\}.
\end{equation}
\end{widetext}
The limit angles $\Theta _{_1 }^{\mbox{left}} ,\Theta_{2}^{\mbox{left}} $
for the left wing are found in the form
\begin{widetext}
\begin{equation}
\label{eq6}
\Theta _{1}^{\mbox{left}} =\pi -\mbox{arcos}\left\{ {-\frac{R+r+a\sin \varphi 
-\Delta x\cos \varphi -2R\cos ^2\varphi }{\sqrt {\left[ {a+(R+r)\sin \varphi } 
\right]^2+\left[ {\Delta x+(R-r)\cos \varphi } \right]^2} }} \right\},
\end{equation},
\begin{equation}
\label{eq7}
\Theta _{_2 }^{\mbox{left}} =\mbox{arccos}\left\{ {-\frac{r+a\sin \varphi 
-\Delta x\cos \varphi }{\sqrt {a^2+\Delta x^2+r^2+2a\,r\sin \varphi -2\Delta 
xr\cos \varphi } }} \right\}.
\end{equation}
\end{widetext}
\begin{figure*}[htbp]
\hypertarget{fig3}
\centerline{
\includegraphics[width=2.0in,height=2.0in]{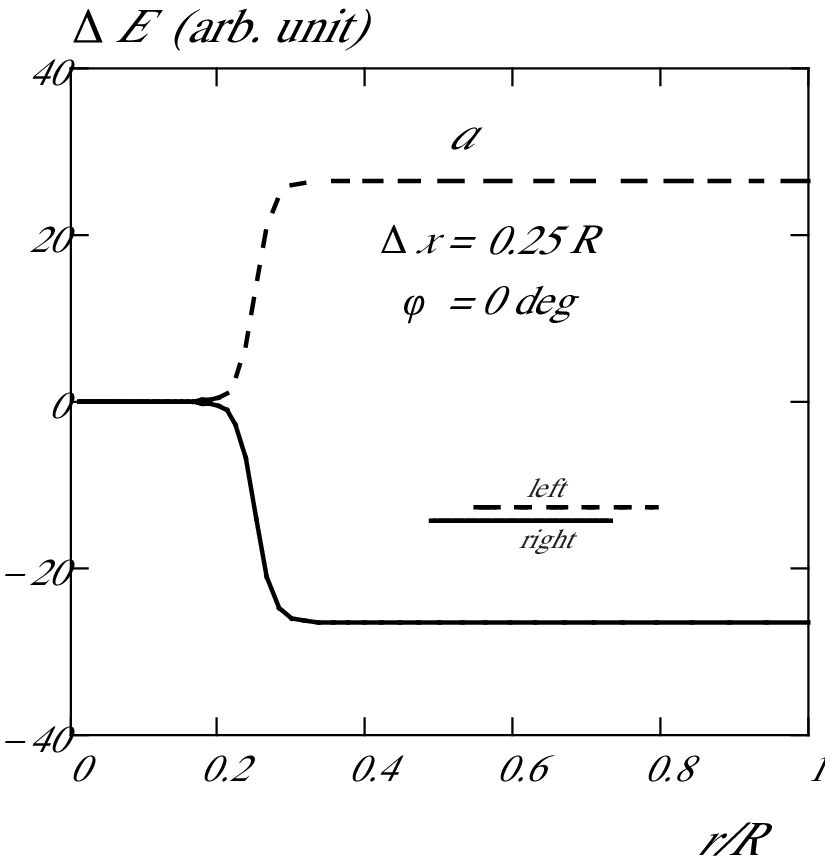}
\includegraphics[width=2.0in,height=2.0in]{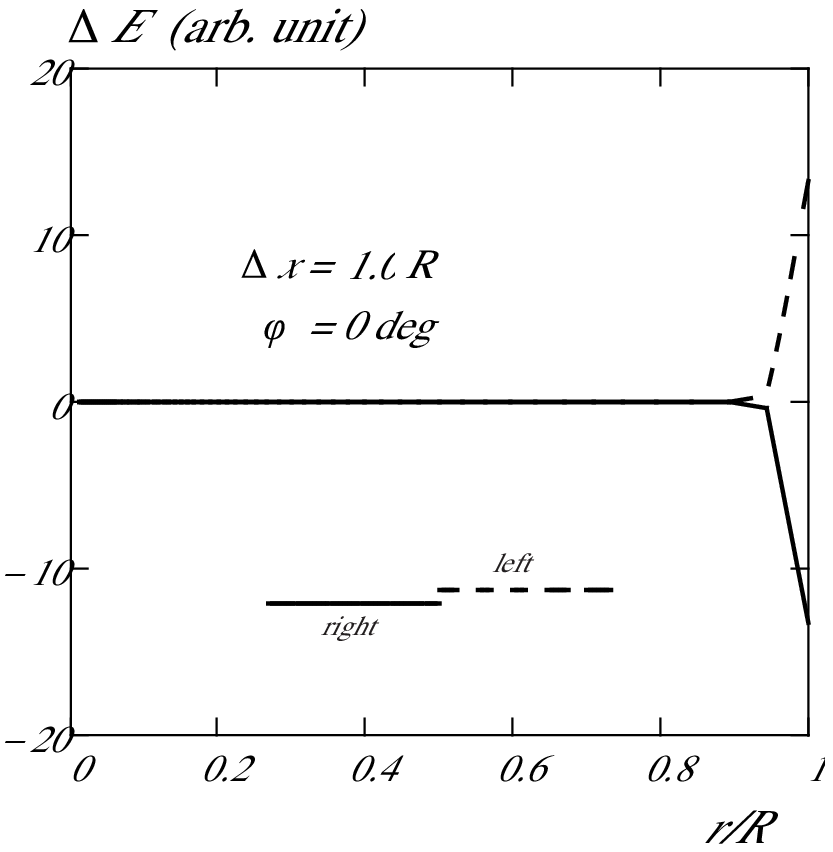}
\includegraphics[width=2.0in,height=2.0in]{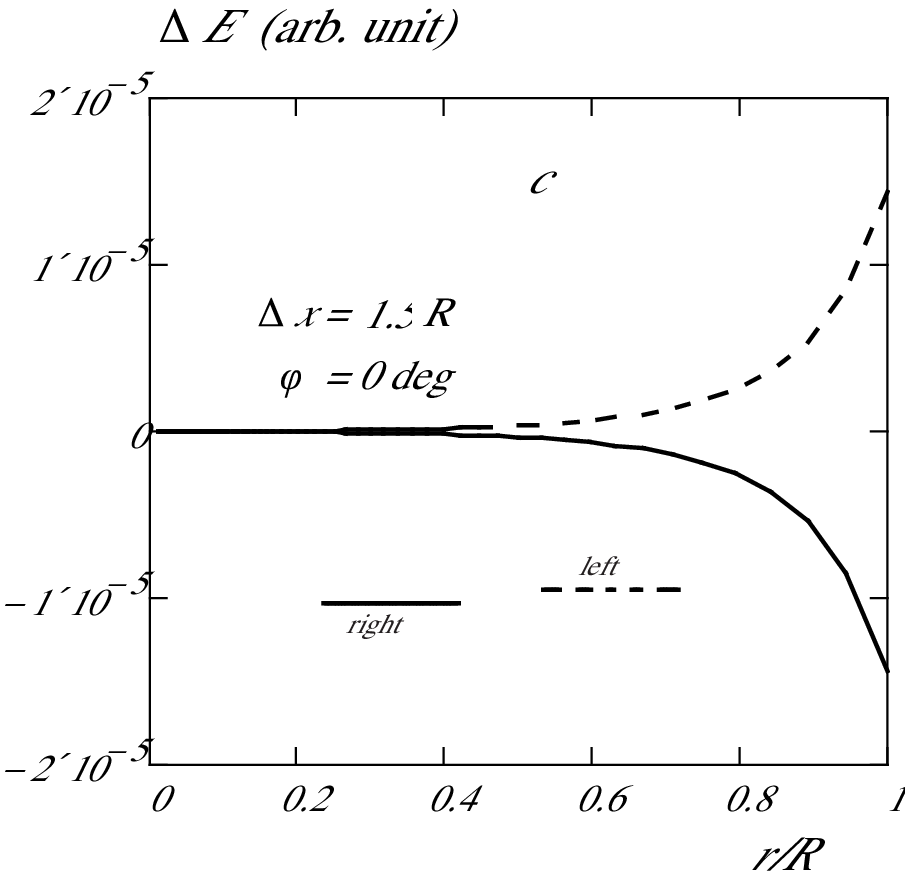}
\label{fig3}}
\caption{Integral values of the EMF $\Delta E$ generated along the parallel 
wings of the figure ($\varphi =0)$ at three different values of the relative 
shift $\Delta x/R$ for the same value of the distance $a$.}
\end{figure*}
\begin{figure*}[htbp]
\hypertarget{fig4}
\centerline{
\includegraphics[width=2.0in,height=2.0in]{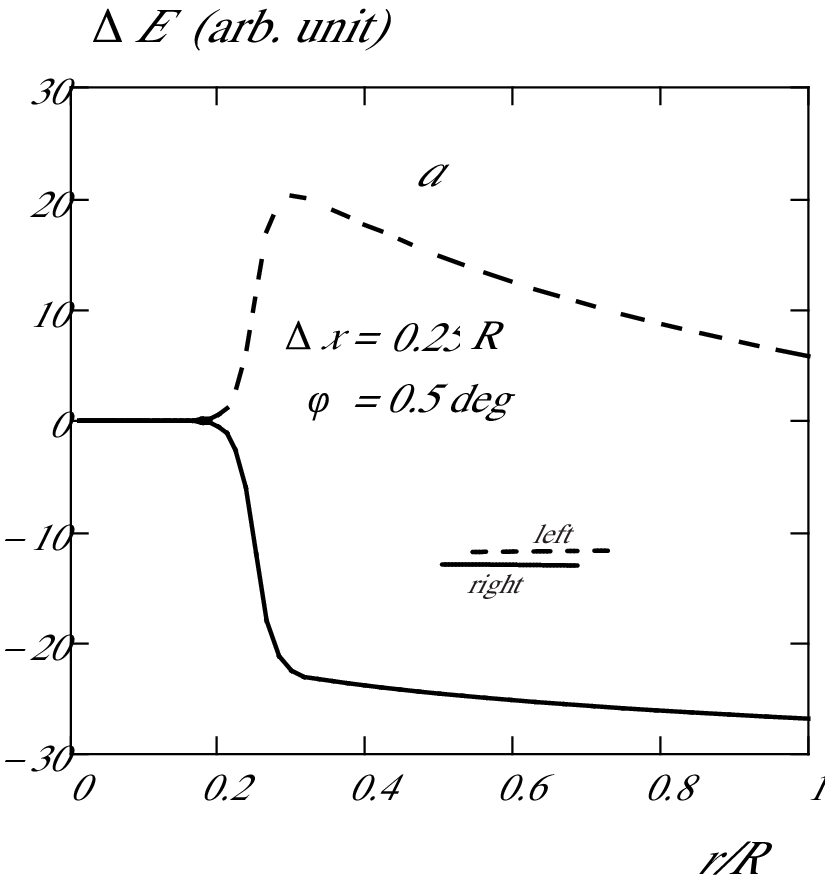}
\includegraphics[width=2.0in,height=2.0in]{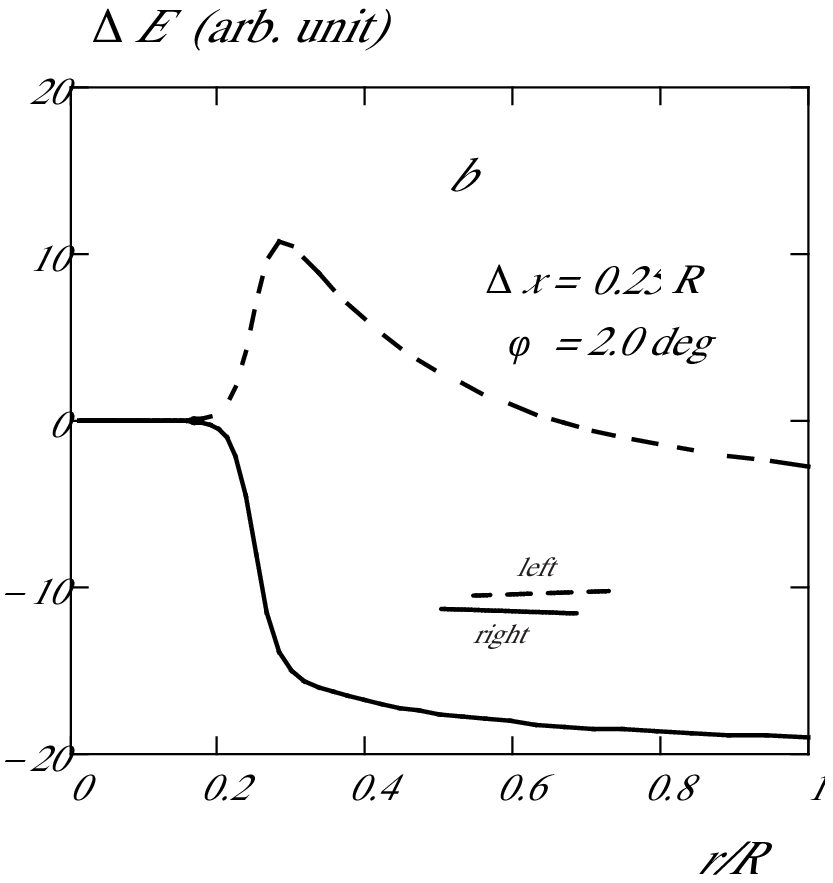}
\includegraphics[width=2.0in,height=2.0in]{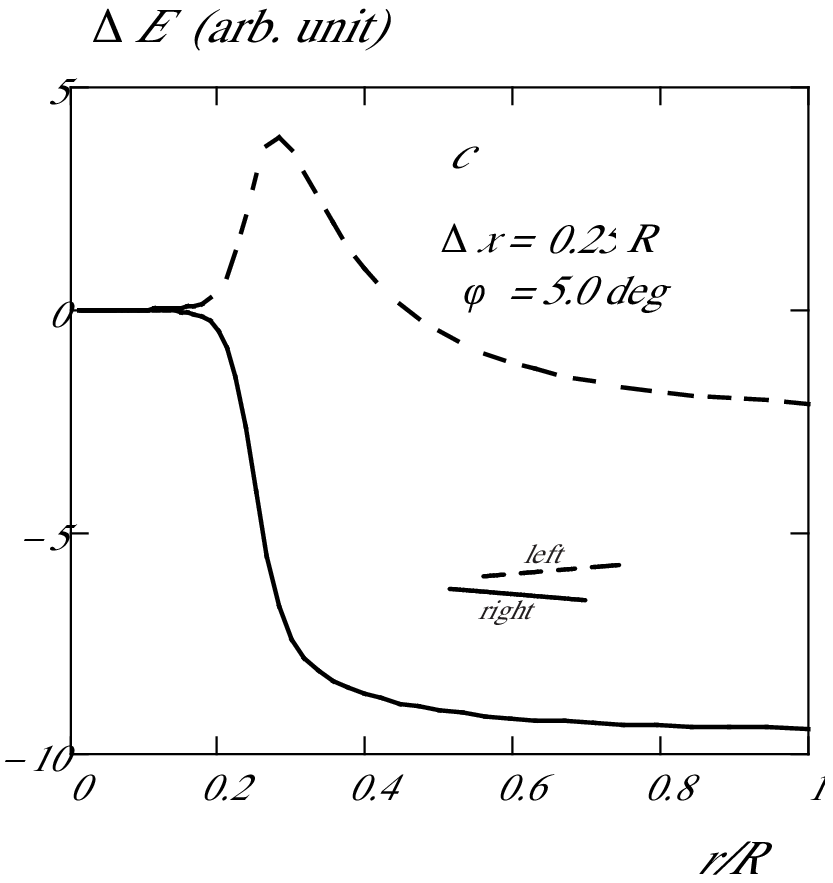}
\label{fig4}}
\caption{Integral values of the EMF $\Delta E$ generated along the figure 
nonparallel wings ($\varphi \ne 0)$ at three different values of the relative 
shift $\Delta x/R$ for the same value of the distance $a$.}
\end{figure*}
In this case, the parameter $s$ in formula (\ref{eq2}) will correspond to the right 
and left wings in the form
\[
s^{\mbox{right}}=\frac{\sin (2\varphi -\Theta _{_2 }^{\mbox{right}} )(a+r\sin 
\varphi )}{\sin (\varphi -\Theta _{_2 }^{\mbox{right}} )} ,
\]
\begin{equation}
\label{eq8}
s^{\mbox{left}}=\frac{\sin (2\varphi -\Theta _{_2 }^{\mbox{left}} )(a+r\sin 
\varphi )}{\sin (\varphi -\Theta _{_2 }^{\mbox{left}} )}.
\end{equation}
Thus, here the basic diagram is presented for calculating the EMF 
generation in optical approximation in nanosized metal configurations due to 
virtual photons for the elements shifted relative to one another.

\textbf{CALCULATION RESULTS}

Using formulae (\ref{eq1}-\ref{eq7}), it is possible to find the following character of 
the EMF generation at the relative shift of two parallel plates which 
are not closed in circuit $\varphi =0$ and have the same length $R$; 
the distance between them is $a$ (see Fig.\hyperlink{fig2} 2). Integral electromotive forces 
corresponding to the shifts and generated along the entire length of each 
of the wings are shown in Fig.\hyperlink{fig3} 3. From Fig.\hyperlink{fig2}2a it can be seen that without a 
shift ($\Delta x=0$), at the ends of both wings (the left wing - at $r/R=0$ 
and the right - at $r/R=1$) there are gradients of the electric-field $E$ 
potential absolutely similar in shape and value but oppositely directed. 
Consequently, in such system no EMF should generate. However, at the 
slightest relative shift ($\Delta x>0$) of the wings, the value and shape of 
the field $E$ intensity function nearby the ends have an asymmetrical 
character (Fig.\hyperlink{fig2}2b). As the result, inside each of the wings, the EMF with an 
opposite relative gradient of the potentials should be generated 
(Fig.\hyperlink{fig3}3a,b,c). In principle, the asymmetry of electric potentials should 
appear at any relative shift ($0<\Delta x<\infty $) changing only in shape 
and value (Fig.\hyperlink{fig2}2c,d) but not in the direction of the gradient of the 
potentials for each of the two wings. Correspondingly, the EMF in the plates 
should be generated at any shift (Fig.\hyperlink{fig3}3a,b,c). The same character of the 
dependences will be present at the rescaling of the dimensional parameters 
of the configuration to any small values within physically reasonable limits 
restricted by possible sizes of atoms and interatomic distances in the 
material of the metals plates.

The character of the EMF generation at the relative shift of two nonparallel 
plates ($\varphi \ne 0)$ significantly differs from that in the situation with 
parallel wings ($\varphi =0)$. For the wings with the same length $R$ and the 
minimal distance $a$ between them at the shift $\Delta x/R=0.25$, the following 
EMF behavior can be observed (Fig.\hyperlink{fig4}4). It can be seen that for the same value 
of the shift, at the increase of the angle $\varphi $ and the growth of the 
left wing length $R$, the EMF generation decreases in the wing and the 
direction of the potential gradient changes to the opposite. At the same 
time, the EMF generation in the right wing only increases relative to that 
in the left one; though it decreases in the absolute value, but the 
potential gradient direction does not change.
\begin{figure*}[htbp]
\hypertarget{fig5}
\centerline{
\includegraphics[width=2.0in,height=2.0in]{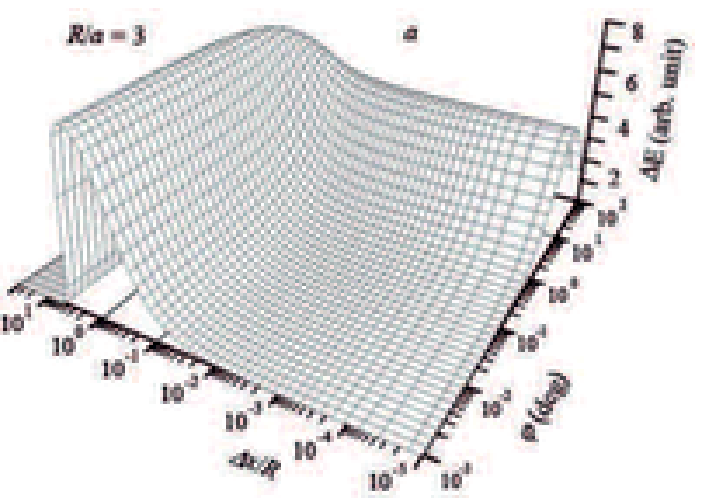}
\includegraphics[width=2.0in,height=2.0in]{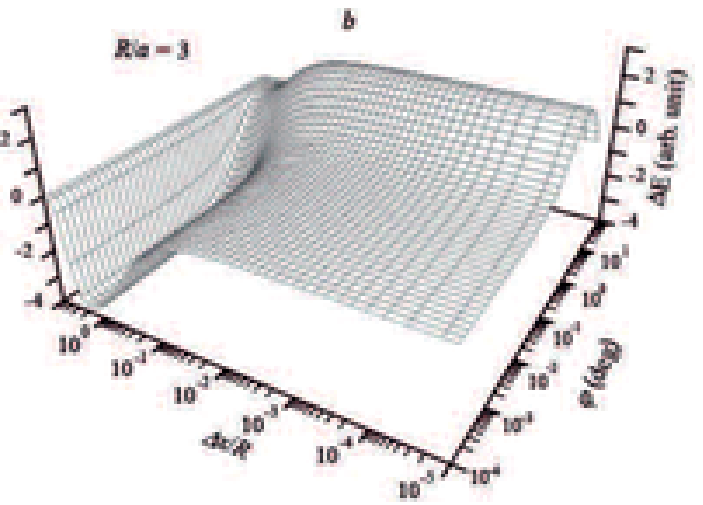}
\includegraphics[width=2.0in,height=2.0in]{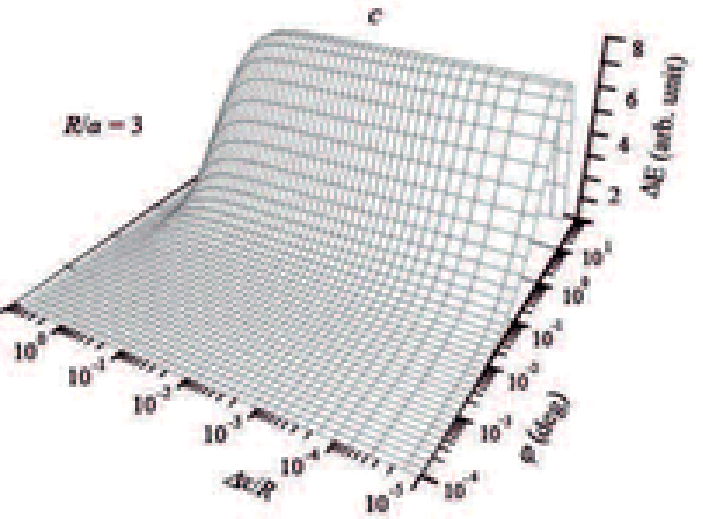}
\label{fig5}}
\caption{Joint EMF dependences on the shifts $\Delta x/R$ and angles $\varphi $ 
at $R/a=3$ for the left ($a)$ and the right ($b)$ wing with the total EMF (c) when 
they are connected in series in circuit. }
\end{figure*}
\begin{figure*}[htbp]
\hypertarget{fig6}
\centerline{
\includegraphics[width=6.0in,height=2.0in]{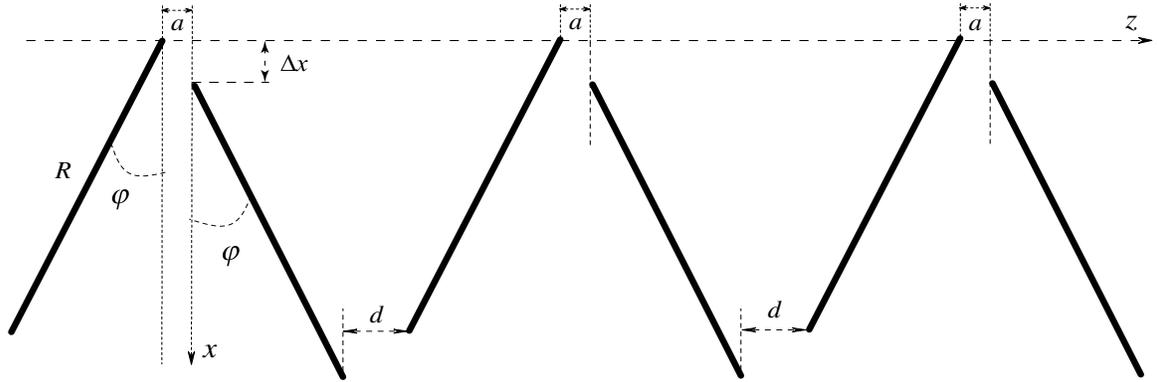}}
\caption{A schematic view of the periodic configuration with nonparallel 
figures located along the $z$-axis at the distance $d$ between them with the 
shift $\Delta x$. }
\end{figure*}
The level of the EMF generation in the joint dependence on the shifts ($0\le 
\Delta x/R\le 40)$ and angles $0<\varphi \le 90^{\circ}$ between the plates is 
shown in Fig.\hyperlink{fig5}5a,b,c. In Fig.\hyperlink{fig5}5 it is possible to observe interesting phenomena both 
for the left and for right wing. Firstly, similar to Figs.\hyperlink{fig2}2 and \hyperlink{fig3}3, it can be 
seen that the EMF can be generated only at the shift of the plates relative 
to one another. Secondly, it appears that for both wings the maximal values 
(in the absolute value) of the EMF generation $\Delta E(\Delta x)$ depending 
on the shifts are larger, than the EMF maxima $\Delta E(\varphi )$ depending on 
the angles between the wings. At the shift, the maximum of the EMF 
generation remains at the same level at any angle $0<\varphi \le 
90^{\circ}$ (Fig.\hyperlink{fig5}5a,b).

Let us note that when the two wings are connected in series in circuit, due 
to the shift the total EMF generation in them is completely compensated at 
any of the angle $0<\varphi \le 90^{\circ}$ (Fig.\hyperlink{fig5}5c). In this case, the shift 
$\Delta x/R>1$ leads to the gradual disappearance of the generation of the 
total EMF in the circuit due to the opening of the wings at any angle.

\textbf{PERIODIC CONFIGURATIONS WITH SHIFTED ELEMENTS} 

Further, let us consider the character of the EMF generation in the chains 
of the wings shifted in relation to one another and placed in periodic 
configurations. Let it be the case of nonparallel figures periodically 
placed along the $z$-axis as shown in Fig.\hyperlink{fig6}6. Each of the figures in the 
configuration period is completely similar to the individual figure (see 
Fig.\hyperlink{fig1}1). In the period, between the ends of the figures there is a distance 
$d$. One on the wings is shifted along the $x$-axis by an arbitrary value 
$\Delta x$ in the direction of $x$-axis (one of the wing's ends touches the 
$x$-axis) without a change in the angle $\varphi $

\begin{figure*}[htbp]
\hypertarget{fig7}
\centerline{
\includegraphics[width=2.0in,height=2.0in]{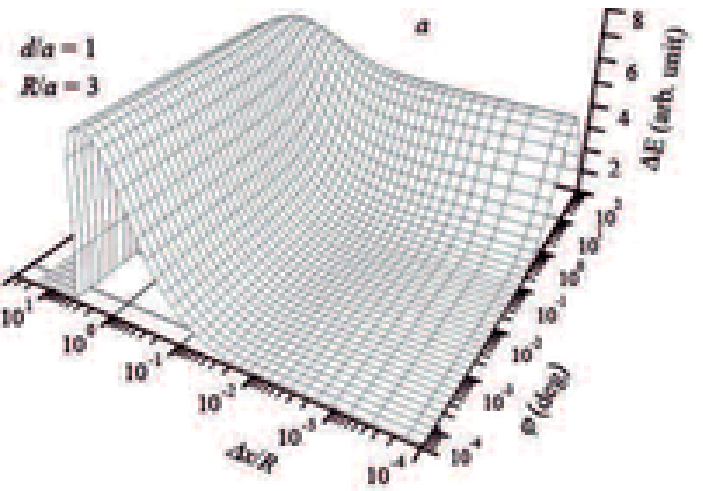}
\includegraphics[width=2.0in,height=2.0in]{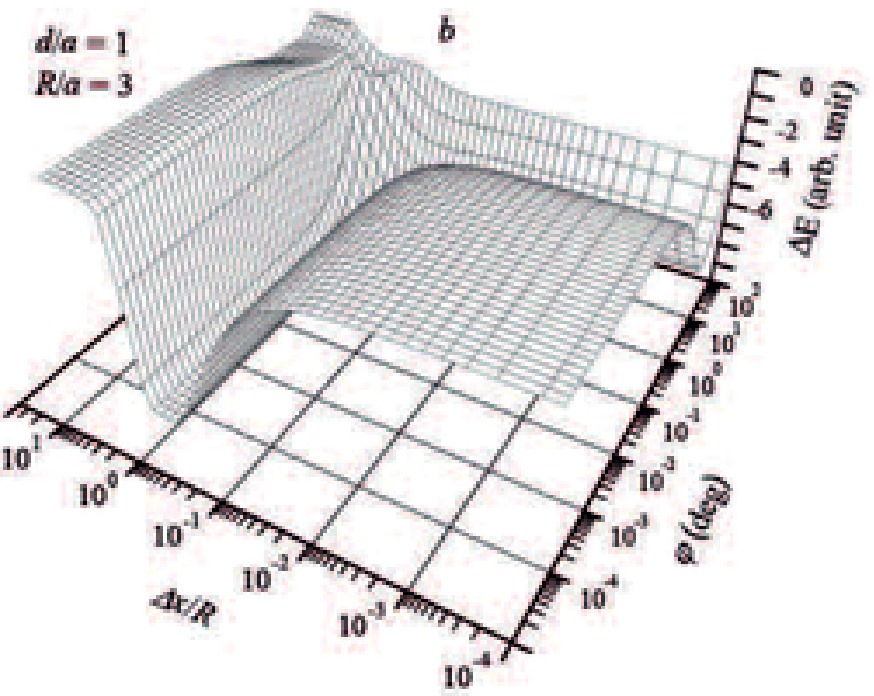}
\includegraphics[width=2.0in,height=2.0in]{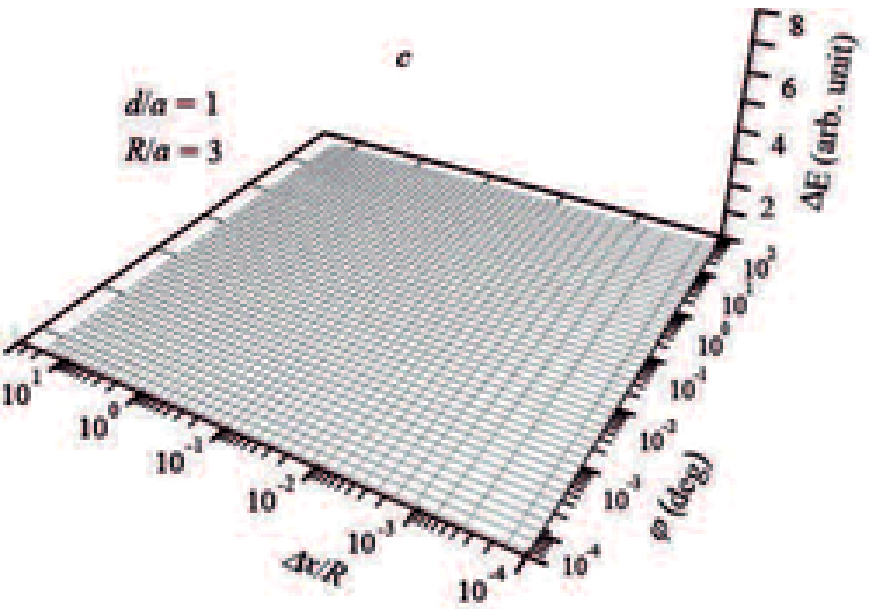}}
\label{fig7}
\caption{Joint dependences of the EMF on the shifts $\Delta x/R$ and angles 
$\varphi $ at $R/a=3$ for two electromotive forces in the right wing: ($a$) -- when 
the gradient of the electric potential is increasing along the $x$-axis; 
($b$) -- when the gradient along the $x$-axis is decreasing; ($c$) -- the sum of 
the two electromotive forces.}
\end{figure*}
\begin{figure*}
\hypertarget{fig8}
\centerline{
\includegraphics[width=2.3in,height=2.3in]{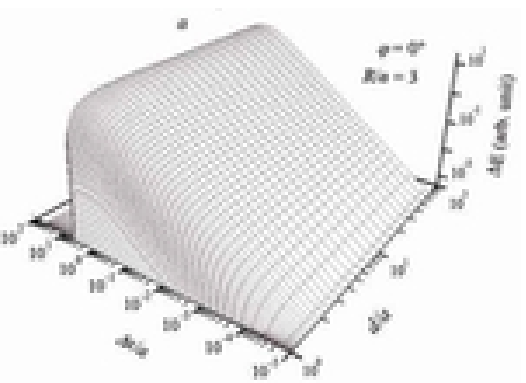}
\includegraphics[width=2.3in,height=2.3in]{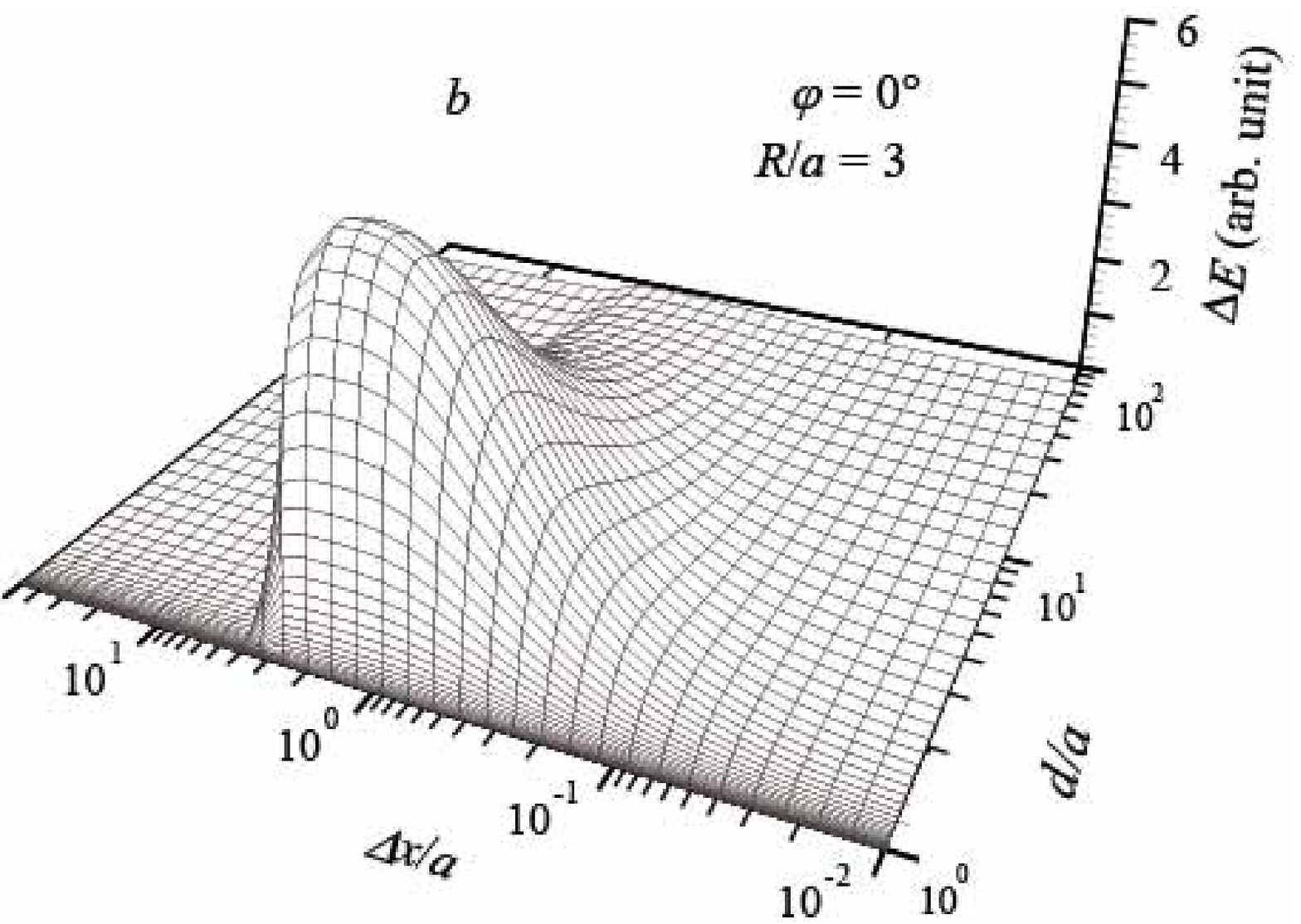}\linebreak 
\includegraphics[width=2.3in,height=2.3in]{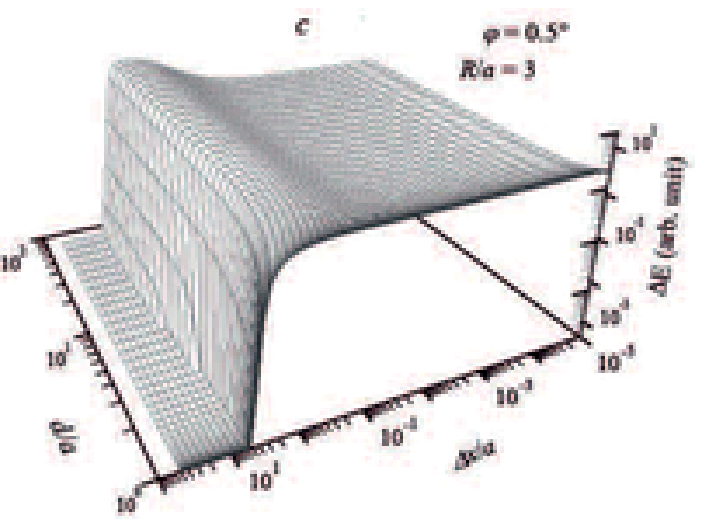}
\includegraphics[width=2.3in,height=2.3in]{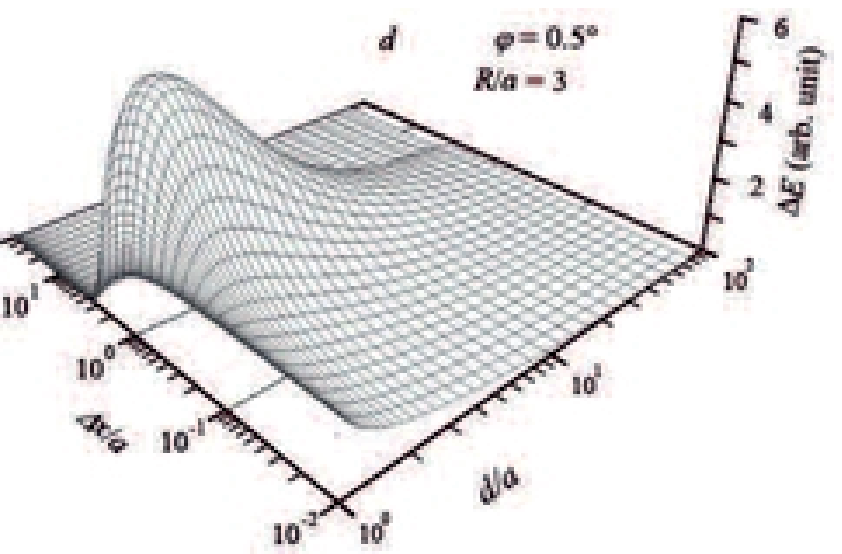}}
 \caption{The EMF (a) and effectiveness $Q$ (b) dependences on the relations 
$d/a$ and shifts $\Delta x/a$ at $\varphi =0^{\circ}$ and $R/a=3$ in the 
periodic configuration, when the wings are connected in series in circuit; 
($c$) and ($d$) similar dependences at $\varphi =0.5^{\circ}$. }
\end{figure*}
\begin{figure*}[htbp]
\hypertarget{fig9}
\centerline{
\includegraphics[width=6.0in,height=4.0in]{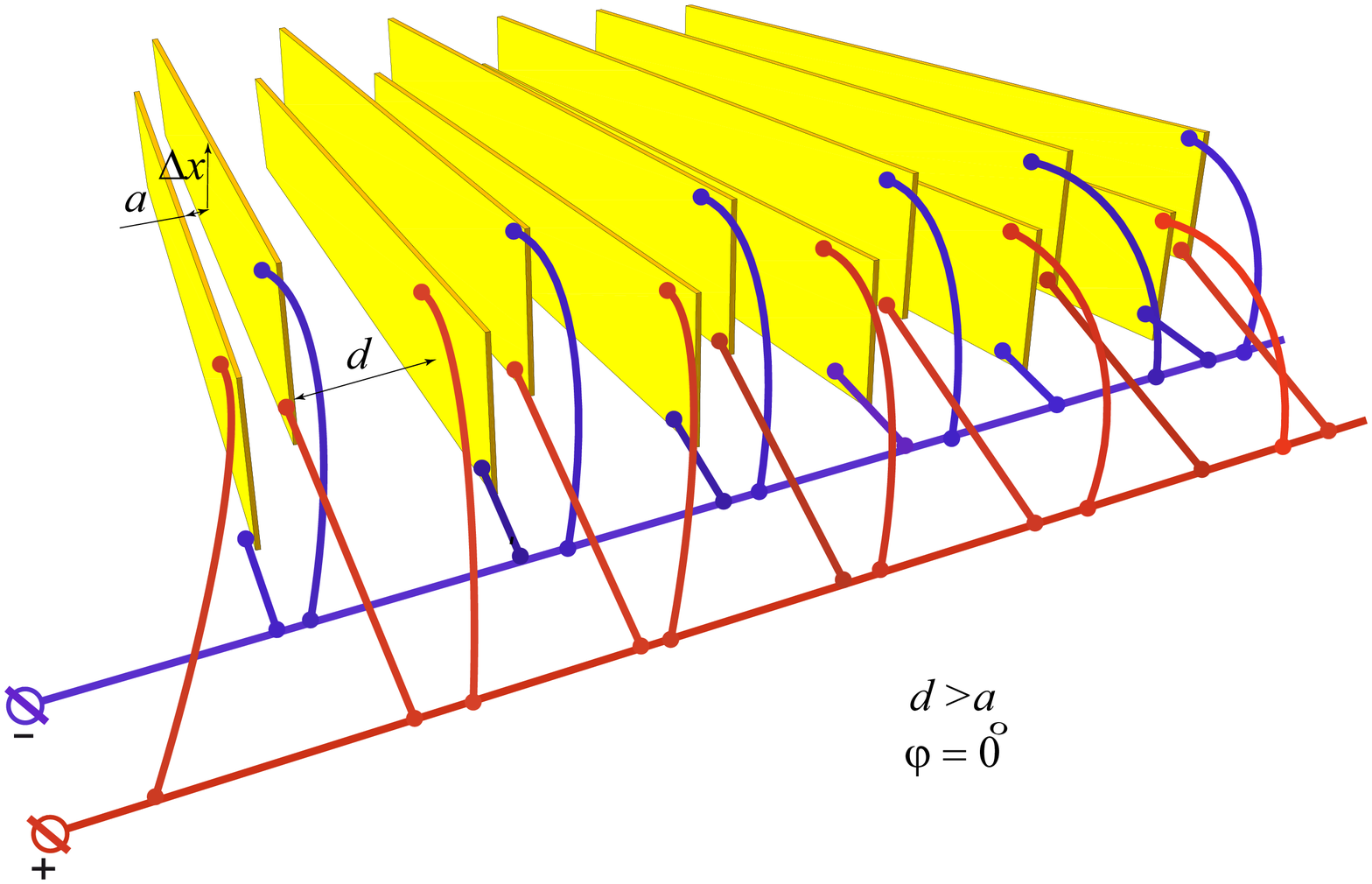}}
\caption{The plot of the Casimir EMF generator in the periodic configuration 
at the parallel connection of the wings in the circuit at the ratio $d/a>1$ 
and shift $\Delta x$ at $\varphi=0^{\circ}$}
\end{figure*}
Similar to the paper on Casimir expulsion forces for periodic configurations 
with nonparallel figures [15], let us note the following. The periodic 
location of $n$ nonparallel figures, the wider opening of which is directed 
against the $x$-axis, with the distance $d$ between them leads to the 
formation of $n-1$ figures with oppositely directed openings (see Fig.\hyperlink{fig2}2). In 
this case, a wing (one of the surfaces of a nonparallel figure) of each of 
the figures is a wing of another figure, the opening of which is directed to 
the opposite side. Thus, it is possible to write an expression for the total 
EMF for $n$ figures periodically located along the $z$-axis 
\cite{Rizzoni:2008} (when the wings are connected in series in circuit 
and not in parallel as sources of an EMF).
\begin{equation}
\label{eq8}
\Delta E_{\sum} =\sum\limits_{n=1}^n {\Delta E_n } =n\Delta E(a)-\left( {n-1} 
\right)\Delta E(d).
\end{equation}
Here, $\Delta E(a)$ is an EMF for the distance $a$ between the nearest ends 
of the figures and $\Delta E(d)$ is an EMF for the distance $d$, 
respectively, instead of $a$ in formulae (1-7). Naturally, when all the 
wings in the periodic configuration with nonparallel figures are connected 
in parallel and not in series in the electric circuit, it is more reasonable 
to calculate the sum of the equivalent summed current in the system.

\textbf{CALCULATION RESULTS FOR PERIODIC CONFIGURATIONS WITH SHIFTED 
ELEMENTS}

From formula (\ref{eq8}) it follows that in the periodic configurations, for $d=a$ 
at any number $n$ of figures connected in series in circuit, Casimir EMF is 
$\Delta E_{\sum} \to \Delta E(a)$. That is, even for $n\to \infty $ at $d=a$ 
the EMF in the periodic configuration is always at the level of the EMF for 
one separate figure. It will happen due to the generation of two 
electromotive forces with opposite and equal gradients of the electric 
potential in each non-extreme wing in the periodic configuration as shown in 
Fig.\hyperlink{fig7}7.

In accordance with formula (\ref{eq8}), at $d\ne a$, the EMF of the periodic 
configuration will depend on the relation $d/a$. At the growth of $d/a$, the 
generation of EMF will tend to the dependence on shifts and angles, 
completely similar to that shown in Fig.\hyperlink{fig7}7a. The electromotive forces in 
dependence on the growth of $d/a$ and shifts $\Delta x/a$ at certain angles 
and relations $R/a$ are shown in Fig.\hyperlink{fig8}8a,c. At the growth of the number $n$ of 
nonparallel figures in the periodic configuration, the character of the 
curves will be similar to that of the curves presented; however, their total 
EMF will grow linearly at the growth of $n$ for any angles $\varphi $.

It is possible to determine the effectiveness $Q$ of the EMF generation for 
$n$ cavities with shifted elements as the relation of $\Delta E_{\sum} $ to the 
entire length of the configuration along the $z$-axis in the following form
\begin{equation}
\label{eq10}
Q=\frac{\Delta E_{\sum} }{n\left[ {a+d+2R\tan (\varphi )} \right]}.
\end{equation}
The dependence of $Q$ on the relations $d/a$ and shifts $\Delta x/a$ is 
shown in Fig.\hyperlink{fig8}8b,d.

From Fig.\hyperlink{fig8} 8a, it can be seen that at any relations $d/a>1$, even for $\varphi 
=0^{\circ}$, at the relative shift of the wings $\Delta x/a$ in the circuit 
with in-series connection, an EMF can generate (see Fig.\hyperlink{fig9} 9). The EMF generation sharply 
decreases at $\Delta x/R>3$. For any length of the wings of the figure $R$ 
with different angles $\varphi $ there is a maximum of the effectiveness $Q$ of 
the generation of the total EMF $\Delta E_{\sum}$. At the EMF calculation in 
the configuration under study, for selected parameters, at $\varphi 
=0^{\circ}$, the optima of the relations $\Delta x/a\approx 1.5$ and 
$d/a\approx 1.8$ are found. In this case, for $R\to \infty $, the joint 
maximum exists at $\Delta x/a\to 2$ and $d/a\approx 1.8$. For nonparallel 
wings, for example, at the angle $\varphi =0.5^{\circ}$ the character of the 
dependences of the EMF and $Q$ changes significantly. In this case, the EMF 
generation is almost at the same level at any shifts (Fig.\hyperlink{fig8} 8c) starting with 
the smallest $\Delta x/R$. However, when the shifts are of the order $\Delta 
x/R>3$, at the angles $\varphi =0.5^{\circ}$ and $\varphi =0^{\circ}$ the EMF 
generation sharply decreases. The effectiveness is also higher at the angle 
$\varphi =0.5^{\circ}$ for the smaller relations $d/a$ and $\Delta x/a$ (Fig. \hyperlink{fig8}
8d).

\textbf{CONCLUSIONS}

Thus, in the present paper, the possibility in principle is shown for the 
existence of Casimir EMF in nanosized parallel metal plates (wings) even at 
the smallest shift of one of them. It is found that the maximal values of 
the EMF generation depending on the shifts of the wings are even larger than 
the EMF maxima depending on the angles between the wings. The direction of 
the gradients of the electric-field potential in two shifted plates is 
opposite; however, it can change when the opening angle between them grows. 
It is found that the EMF generation is also possible in periodic 
configurations with shifted elements. The EMF generation takes place even in 
parallel configuration elements shifted with periodicity. There are optimal 
relations in the geometry of a shift and distances between the plates of the 
figures at which there can be a maximal EMF generation in the periodic 
configuration. 
\begin{acknowledgments}
The author is grateful to T. Bakitskaya for hers helpful
participation in discussions.
\end{acknowledgments}

\end{document}